# Electrode modified domain morphology in ferroelectric capacitors revealed by X-ray microscopy

Megan O. Hill Landberg[1*], Bixin Yan[2], Huaiyu Chen[3], Efe Ipek[2], Morgan Trassin[2], and Jesper Wallentin[3]

[1] MAX IV Laboratory, Lund University, 22100 Lund, Sweden
[2] Department of Materials, ETH Zurich, 8049 Zurich, Switzerland
[3] Synchrotron Radiation Research and NanoLund, Department of Physics, Lund University, 22100 Lund, Sweden
*Corresponding author: megan.landberg@maxiv.lu.se

**Abstract**

Ferroelectric thin films present a powerful platform for next generation computing and memory applications. However, domain morphology and dynamics in buried ferroelectric stacks have remained underexplored, despite the importance for real device performance. Here, nanoprobe X-ray diffraction (nano-XRD) is used to image ferroelectric domains inside $BiFeO_3$-based capacitors, revealing striking differences from bare films such as local disorder in domain architecture and partial polarization reorientation. We demonstrate sensitivity to ferroelectric reversal in poled capacitors, revealing expansive/compressive (001) strain for up-/down-polarization using nano-XRD. We observe quantitative and qualitative differences between poling by piezoresponse force microscopy (PFM) and in devices. Further, biasing induces lattice tilt at electrode edges which may modify performance in down-scaled devices. Direct comparison with PFM polarized structures even demonstrates potential nano-XRD sensitivity to domain walls. Our results establish nano-XRD as a noninvasive probe of buried ferroelectric domain morphologies and dynamics, opening avenues for operando characterization of energy-efficient nanoscale devices.

## Introduction

The study of ferroelectric thin films has been revived by the recent push for beyond-CMOS applications and low energy consuming emerging computing schemes.[1,2] The device properties are fundamentally related to the nano to microscale ferroelectric domain architecture. The elastic and electrostatic energy landscape at domain boundaries is impacted by device integration and scaling down, key for application implementation. Most importantly it may evolve upon electrical poling and drive switching properties and hence, final device reliability and functionality. The imaging of ferroelectric domain states, directly in application-relevant metal-ferroelectric-metal capacitor heterostructures, is therefore highly desired to shed light on structure-property relationships in such materials.[3]

Imaging ferroic domains in the buried configuration with high resolution has remained challenging using conventional techniques such as scanning probe, electron or optical microscopy, since the top electrode of capacitors cover the ferroelectric film.[4,5] Therefore, the impact of device integration of ferroelectric materials on domain architecture and the study of their dynamics upon electrical poling at the nanoscale has remained largely unexplored.

Recent developments in X-ray sources and optics have allowed nanoprobe X-ray diffraction (nano-XRD) to reach below 100 nm spatial resolution, while keeping the advantage of long penetration depth. Hruszkewycz et al. imaged the dissolution of vertical DWs under in-plane electric fields,[6] and recently Guzelturk et al. used similar methods to image domain evolution under photoexcitation.[7] Others have investigated buried ferroelectric films, even studying application of bias, but experimental constraints prevented researchers from spatially resolving domain structures.[8,9]

Here, we use nano-XRD to image nanoscale domains within a ferroelectric capacitor structure, providing newfound clarity into ferroelectric domain ordering under metallic electrodes and upon polarization switching. Taking ferroelectric $BiFeO_3$ (BFO) as our model system,[10] we resolve nanoscale domains directly within capacitors and demonstrate X-ray sensitivity to polarization switching in both bare films and under thick electrodes. We reveal that mere deposition of metal contacts induces a domain pattern modification, prior to any applied electric field application. In capacitors, we also observe strikingly different domain response to the applied electric field around the edge of the contacts, pointing to notable effects in polarization switching upon downscaling. While revealing unexpected structure changes in $BiFeO_3$ devices, this work also highlights the potential for nano-XRD as a powerful tool to investigate a wide range of ferroelectric devices.

**Results**

*Domain disorder under electrodes*

We first studied the domain pattern of as-deposited ferroelectric thin films, comparing nano-XRD and conventional piezoresponse force microscopy (PFM). Motivated by the recent report on ultralow energy consuming logic devices based on the integration of ferroelectric magnetoelectric BFO, we grew our BFO on $SrRuO_3$-buffered (SRO) $DyScO_3$ (DSO).[11] Rhombohedral (R3c) BFO presents four ferroelastic domain variants with polarization pointing along one of four <111> pseudo-cubic (pc) directions, with eight total ferroelectric polarization directions. The ~0.4% compressive epitaxial strain results in a ferroelectric domain architecture primarily consisting of stripe-like 71° ferroelastic domains oriented along DSO [001].[12,13] Top platinum electrodes were then deposited on the BFO/SRO/DSO films to finalize the metal-ferroelectric-metal capacitor heterostructures. The schematic in Figure 1a shows the striped domain structure with respect to DSO and BFO crystallographic axes. Nano-XRD was performed in the geometry shown in Figure 1b, collecting high resolution 2D diffraction patterns of the (003)pc peak, as exampled in Figure 1c, at each probe position. The center of mass of the Bragg peak was calculated for each probe position, producing tilt and strain maps (see Methods).

We first show results from an unpoled BFO capacitor structure with a Pt top electrode ("e:0V"). The comparison between PFM and nano-XRD demonstrates that the nanoscale BFO domain architecture, here ~100 nm wide stripe-like 71° ferroelastic domains can be spatially resolved using both techniques. The superiority of the nano-XRD approach however strikingly appears when addressing the nanoscale domain structure buried under 100 nanometers of Pt. While the thick metal electrode screens the AC-bias necessary for PFM readout, nano-XRD imaging remains unaffected, as shown in the side-by-side comparison in Figure 1d,e. Nano-XRD can generate several types of contrast, further discussed below, and here we show the $\alpha$-tilt which refers to rotation around $q_z$ (defined in 1b). This is approximately tilt of the (001) planes *between* ferroelectric domain stripes (along $[100]_{PC}$), as schematized in Figure 1f. Indeed, a ferroelastic tilt is expected between 71° domain stripes. To allow for statistical analysis, a larger map was taken in the vicinity of another contacted BFO region (Supplementary Note 2). 2D FFT analysis generates log power spectral density maps (Figure 1f) for a 135 µm² area from an as-grown (AG) region (left) and a region buried under the electrode (right). The AG region shows clear vertical lobes at about ± 5 µm$^{-1}$, indicative of highly aligned horizontal stripes domains with ~100 nm domain width. The contacted region on the other hand shows a much more diffuse FFT, pointing to more irregular domain structure. Due to this disorder, the distribution of tilts is also slightly lower for the contacted region (Figure S3d).

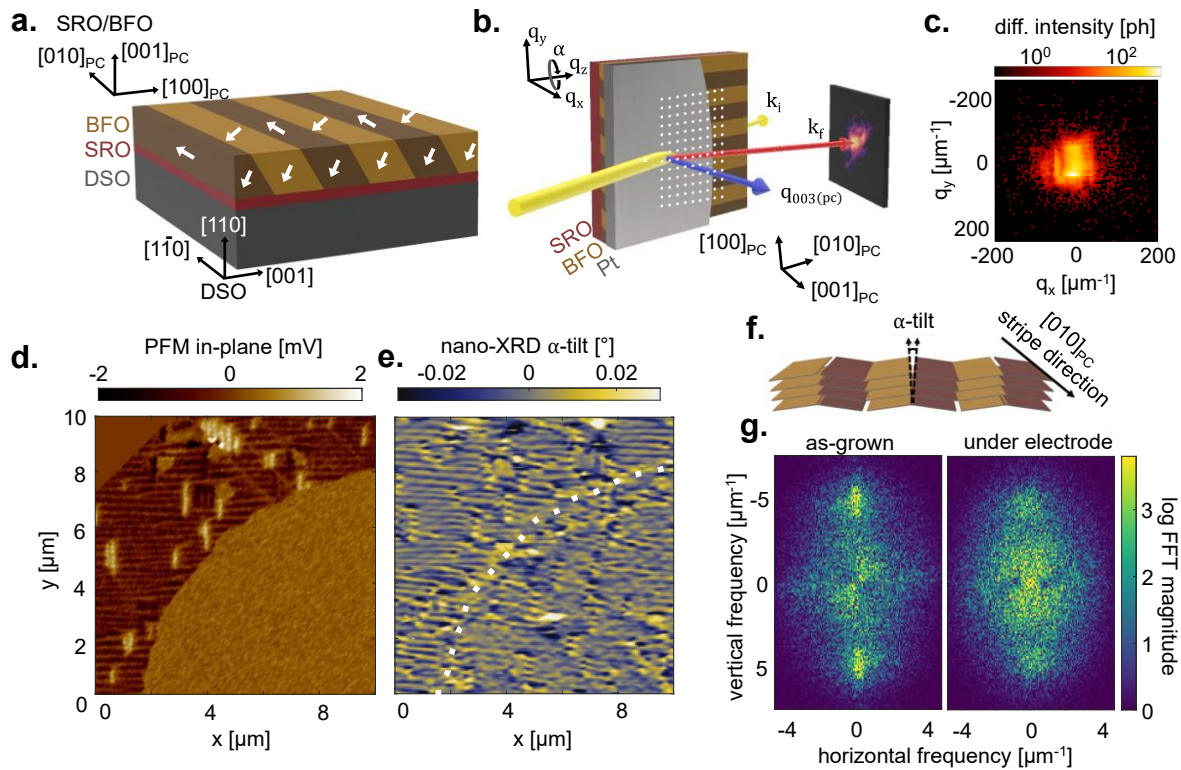

**Figure 1:** (a) Schematic of BFO stack with AG ferroelectric domain directions marked in white. Axes given for the DSO crystallographic directions and the SRO/BFO pc directions. (b) Nano-XRD scattering geometry for the capacitor stack with the X-ray probe in yellow, *q*-vector in blue, and $k_f$ in red. Tilt direction $\alpha$ described by the rotation around $q_z$.

Example probe positions marked in white. (c) Exemplary diffraction pattern of BFO (003)pc for AG region of the film (d) In-plane PFM map of unpoled electrode e:0V (~1/4 of electrode shown). (e) Nano-XRD α-tilt map for the same electrode. Dotted white line shows the outline of the contact extracted from PFM. (f) Schematic describing the tilt relationship of α with the domain stripe direction. (g) 2D FFT magnitude of AG (left) and contacted (right) regions of BFO from Figure S3b.

*Imaging ferroelectric reversal in capacitors*

Next, nano-XRD was used to probe polarization reversal in electrically switched capacitors. Three electrodes were biased to induce positive polarizations with out-of-plane polarization component towards the Pt electrode. Notably, all up-polarized (p-up) capacitors show a (003) lattice expansion compared to down-polarized (p-down) AG regions and unpoled electrode. Further, prominent $\alpha$-tilt variations are observed along the contact edge in poled contacts. We observe measurable strain and tilt variations in the capacitors, despite poling to produce pure 180° ferroelectric switching, suggesting that ferroelectric polarization reversal in BFO is coupled to out-of-plane structure changes.

Figure 2 compares poled e:6V (p-up) with pristine device e:0V (p-down) using three different types of contrast taken from the same nano-XRD maps. The α-tilt images reveal a stark and surprising difference in domain response between the edge and the device interior (Figure 2a,b). In the device interior we see little significant change due to biasing. Indeed, when plotting an α-tilt histogram (Figure 2e-top), there is no variation across the four electrodes, indicating that the average domain structure does not change significantly upon biasing. In contrast to the average tilt distribution, we observe a clear increase in α-tilt around the contact edge of biased device (e:6V) shown in Figure 2a-bottom. In comparison, this edge effect is present but barely visible in pristine device (e:0V), Figure 2a-top. The edge effect should be considered as a potentially significant contribution to device behavior, as strain and tilt are closely intertwined with polarization. A careful comparison shows that the affected region extends from the contact edge to about 0.75 µm inside, making it invisible to PFM. In our large prototype devices (20 µm diameter) the high-tilt ring is already ~14% of the device area. However, for more typical device scales, for example a 1 µm radius, this high-tilt ring would make up >90% of the overall area, dominating device behavior.

Figure 2c shows diffraction intensity for e:0V (top) and e:6V (bottom). While there are local variation in diffraction intensity from disorganized domain regions, there is no significant difference in intensity of BFO under the electrodes versus in the AG regions for either biased or pristine devices. This invariance in diffraction intensity means that (003) strain can be extracted from a single diffraction angle (see Supplementary Notes 1,3). Figure 2d shows $\varepsilon_{003}$ strain for e:0V (top) and e:6V (bottom). Comparing the global behavior of strain under electrodes versus AG regions, we see mostly invariant strain for e:0V. However, e:6V shows a measurable increase in strain under the electrode as compared to the AG region. This suggests that biasing the capacitor to the p-up condition induces an expansion of the out-of-plane lattice parameter. Indeed this is seen for all biased electrodes, as shown by the histogram in Figure 2e-bottom.

Here, the $\varepsilon_{003}$ strain distribution, normalized by the surrounding AG region, is shown for all four contacts. While the mean strain is different for each electrode, all biased electrodes (e:6V, e:8V, e2:6V) show an increase in strain. The mean strain difference between the p-down (e:0V) and p-up (e:6V) is 0.025%. Nano-XRD maps for all other devices are in Supplementary Note 4.

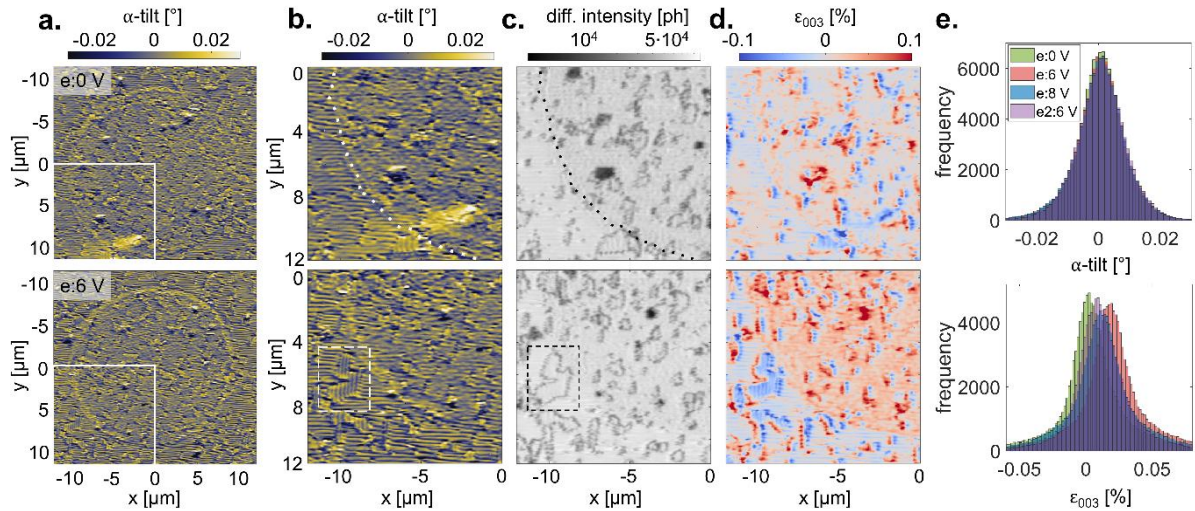

**Figure 2:** (a) Nano-XRD maps of α-tilt for e:0V on top and e:6V on bottom. The region marked with a white square is extracted in (c-d) (b) Zoomed in maps of (b) α-tilt, (c) diffraction intensity, and (d) estimated $\varepsilon_{003}$ strain for e:0V (top) and e:6V (bottom). Dotted white/black lines show the outline of the contact pad extracted from PFM maps. Dotted lines are not shown for maps where it would disrupt contrast. Dashed white and black boxes highlight an example region where vertical and horizontal domain stripes meet (109° domain wall). (e) α-tilt distribution (top) and strain distribution (bottom) histograms for all positions under four electrodes: e:0V is pristine (p-down) and e:6V, e:8V, and e2:6V are switched (p-up).

*Interference effects from domain walls*

Both under electrodes and in the AG film, we observe a mix of horizontal and vertical domain structures. Exampled by the area within the dashed box in Figure 2b,c, the regions in which vertical and horizontal domains meet constitute 109° DWs. Interestingly, at this interface, we observe a significant decrease in the diffraction intensity (Figure 2c) which is not seen at the 71° domain walls (see Supplementary Note 5). The 109° DWs are parallel to the scattering vector (along [001]$_{pc}$) which may produce more pronounced sensitivity to structural changes at DWs, despite their small size (1-5 nm). For instance, regardless of DW type (Ising, Bloch, Néel),[14] out-of-plane polarization will be zero at the DW, inducing localized electrostriction in the [001]pc direction. It has also been previously observed that even purely ferroelectric (180°) DWs are mechanically unique from their surroundings[12] Though DWs are not resolvable with the 60 nm probe, such features can produce interference effects that reduce the scattering intensity. Given this observed sensitivity, it is likely that X-ray nanoprobes will be useful for investigating the structure of DWs in ferroelectrics.

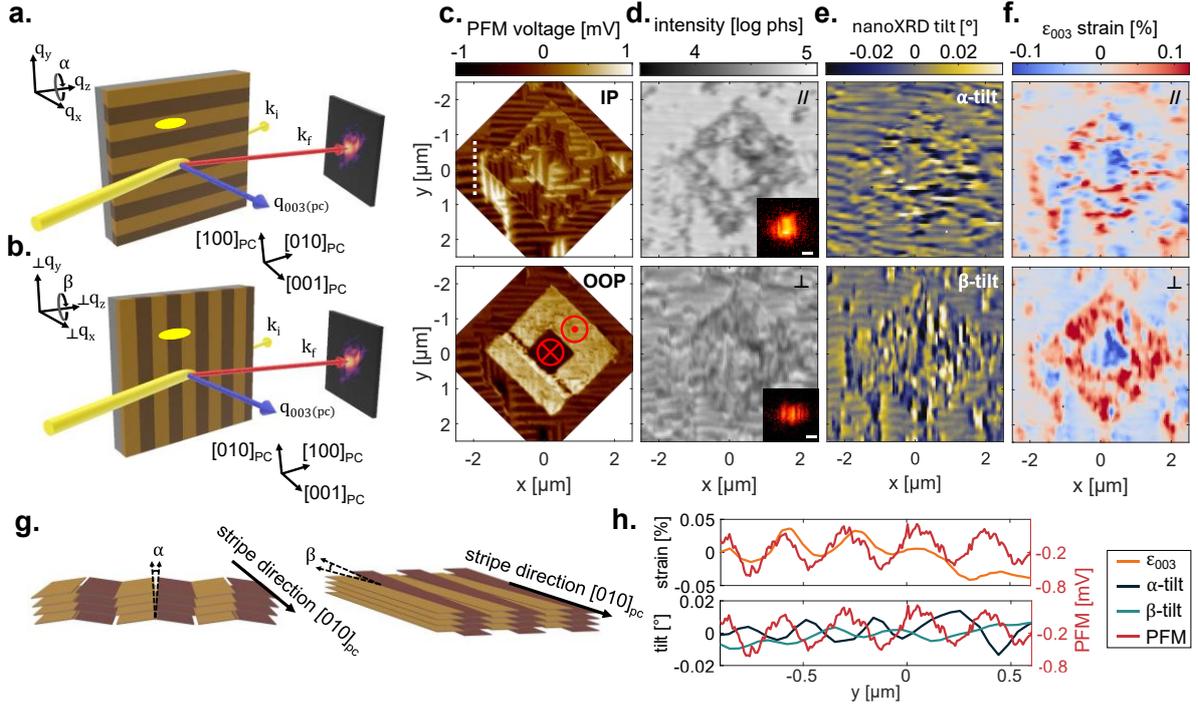

**Figure 3:** Scattering geometry for the probe parallel (a) and perpendicular (b) to the domains. X-ray probe in yellow, q-vector in blue, and $k_f$ in red. Approximate beam footprint with respect to the domains is shown by a yellow ellipse. (c) PFM maps of in-plane (top) and out-of-plane (bottom) polarization within the box-in-box region of the BFO, written by PFM. Red markers indicate polarization directions which were induced by PFM poling of +/- 5V. (d) Diffraction intensity maps from parallel (top) and perpendicular (bottom) scattering of the same box-in-box region as (c). Insets shows exemplary diffraction pattern from the AG regions. Scale bars are 25 µm$^{-1}$. (e) Nano-XRD tilt maps for α-tilt and β-tilt, defined by rotation around $q_z$ for parallel (a) and perpendicular (b) scattering respectively. (f) Strain ($\varepsilon_{003}$) maps from parallel (top) and perpendicular (bottom) scattering. (g) Schematic describing the tilt relationship α- and β-tilt with respect to the domain stripe direction. (h) Line cut extracted from the white dashed line in (c) comparing the in-plane PFM voltage with $\varepsilon_{003}$ (top) and α-/β-tilt (bottom).

*Contrast to ferroelectric switching by PFM poling*

Finally, for more detailed comparison of the sensitivity of nano-XRD to polarization switching, we imaged the bare film in a region around a box-in-box structure written by PFM. Figure 3c shows the in-plane (IP) and out-of-plane (OOP) PFM measurements for the box-in-box region. Here the OOP (bottom) shows expected polarization switching, where the outer box was poled to -5V to produce a positive OOP polarization followed by poling the inner box to +5V to switch back to negative OOP, these direction are indicated by red arrows. The IP PFM (top) shows variation in in-plane polarization between domain stripes. Here we can observe a variety of domain features: 71° domain switching comparable to the AG region, 109° domain switching where the domain stripes meet at 90° in the PFM image, and 180° (purely ferroelectric) switching where the domain stripes change polarization contrast from dark to light.

Diffraction measurements in this region were taken in two geometries, because these were identified to provide different contrast: (1) beam parallel to the ferroelastic domain stripes (as in Figure 3a) and (2) sample rotated 90° such that the beam is perpendicular to the domain stripes (Figure 3b). The beam size is illustrated by yellow ellipses. Overall, nano-XRD produces contrast comparable to that measured by PFM, even though the methods are of an entirely different nature.

We first compare the simplest contrast type, namely diffraction intensity. Figure 3d shows parallel (top) and perpendicular (bottom) integrated diffraction intensity maps for the box-in-box region. The insets depict example diffraction patterns, where a notable difference is the interference fringes in the perpendicular geometry. The fringes in the bottom pattern are spaced approximately 30 µm$^{-1}$ in $q_x$, indicative of a repeating feature of ~210 nm, consistent with ferroelastic stripes of width 105 nm. Due to this interference, the overall scattering intensity of the perpendicular geometry (bottom) is lower than that of parallel scattering (top), even for the same region. This measurement is similar to those in which multiple domains were probed simultaneously.

The box-in-box region is easily identifiable in the diffraction intensity (parallel scattering), with lower intensity in the p-up region. Under first consideration, this would suggest that in the p-up region the Bragg angle is not met. However, taking a rocking curve in this region (Supplementary Note 5, Figure S6) shows that this reduction in intensity is not due to a change in the Bragg condition, but is instead from interference, similar to the interference effects present in the perpendicular geometry that reduce the scattering intensity. This may be due to interference from a high density of 109° and 180° domain boundaries, as discussed previously. In the perpendicular scattering geometry, the intensity variation in the p-up region is less pronounced, possibly because the beam is illuminating multiple DWs regions, averaging out their effect. It then follows that we do not see these interference effects in the poled capacitors as the density of DWs would be significantly lower (only at electrode edges).

As the Bragg condition is not significantly changing between poled regions (Supplementary Note 3), we can estimate tilt and strain within these areas. Figure 3e shows α-tilt (top) and β-tilt (bottom) which are defined by rotation around $q_z$ for parallel and perpendicular geometries respectively. As schematized in Figure 3g, α-tilt probes tilt *between* ferroelectric domains (along [100]pc) and β-tilt probes tilt *along* domains (along [010]pc). In the AG region ferroelectric domains are resolvable, with better resolution for α-tilt compared to β-tilt due to the smaller beam footprint. Within the poled region, given the interference effects, tilt is hard to quantify, but a striped structure matching PFM is still visible.

Figure 3f shows the strain for the parallel (top) and perpendicular (bottom) geometries. These maps are sensitive to the same feature, $\varepsilon_{003}$ strain, but the parallel geometry has a higher spatial resolution along the stripes due to the beam footprint. In both cases, we observe strain contrast from the box-in-box region. This is more prominent for the perpendicular geometry due to the averaging effect across multiple domains, resulting in the largest contrast for polarization changes. Thus, we find that the most striking effect measured with nano-XRD is an out-of-plane strain, with compression for p-down (-0.06%) and expansion for p-up (+0.06%). This effect is similar to the observation for poled capacitors, though the average strain magnitude is higher for PFM poling compared to electrode poling which exhibit only a 0.025% strain difference between p-down and p-up electrodes.

It is difficult to determine the precise nature of the strain effect observed here. A mixture of effects is expected given the presence of multiple domain types within the PFM poled region. The high density of 109° and 180° DWs is likely to produce strain variations similar to those in the AG region where 109° DWs are present (where vertical and horizontal domain stripes meet). These domains could produce higher strain values measured in Figure 3f, giving artificially higher an average strain change with polarization switching as compared to poled capacitors.

For direct correlation with polarization from PFM, Figure 3h shows a line cut extracted from the white dashed line in Figure 3c. Here we observe that strain varies with the same period as the domain stripes, which is unsurprising given the lattice mismatch present due to ferroelastic domain (71°) interfaces. The tilt shows the same trend, both α-tilt and β-tilt. The α-tilt oscillates in the same period as the polarization change. However, interestingly, β-tilt, which is sensitive to tilt *along* domains, is offset about half a period from the domain stripes. While α-tilt is sensitive to ferroelastic tilt, it is possible that this offset in β-tilt could be related to the know chiral structure at the 71° DWs in $BiFeO_3$.[15,16] Further maps of the AG region are shown in Supplementary Note 1 where α-tilt, β-tilt, and strain are more precisely quantified by analysing angular rocking data.

## Discussion

We establish here that deposition of metal electrodes alters the domain structure in ferroelectric thin films. This makes clear that assumptions about the buried film, taken from imaging AG regions, do not hold, as such changes in the domain structure can modify overall performance. In particular, the observation that overall tilt distribution in buried BFO is lower than in the AG film, may have implications for the foreseen device integration of such multiferroic materials. We attribute this modification to metal stress, which is a well known phenomenon in microelectronics,[17,18] modifying the domains by elastic coupling in BFO.[19]

Further, we show quantitative and qualitative differences in domain dynamics for switching by PFM and electrodes. We compare the domain structure of pristine and biased capacitors: biasing induces a tilt around the contact edge which could significantly impact the behavior of smaller devices. Further, we identify lattice spacing changes upon biasing in both capacitors and PFM-poled films. Strain variations between p-up and p-down appear substantially larger for PFM poled films as compared to the capacitors, likely due to high densities of 180° and 109° domain walls in the box-in-box structure producing artificially higher strain magnitudes. However, the overall trend both in PFM poled and capacitor regions suggests that polarizing BFO film produces an out-of-plane [001] change in d-spacing: expansive for p-up and compressive for p-down. This is a surprising result, as a pure 180° polarization reversal should only lead to a displacement of Bi-atoms. However, it is clearly observed here that polarization switching is coupled to a structural change, resulting in a change in (001) lattice spacing. This could be related to electrostrictive effects from insufficient polarization screening or a previously unidentified structural coupling. Note that this is a very small and localized strain effect on the scale of 0.02%, to which other characterization techniques are insensitive. This structural change warrants additional study, as it could exacerbate polarization fatigue present for 180° switching by making domains further susceptible to pinning under increasing internal stresses.[20,21]

We demonstrate that nano-XRD is highly sensitive to 71° ferroelastic domains. Further, we show that nano-XRD may provide sensitivity to DWs via interference effects despite DWs being smaller than the probe. Quantifying such interference effects may be possible via other X-ray imaging techniques such as Bragg ptychography, pointing to opportunities for X-rays to investigate or even image DWs.

It is clear that polarization reversal in BFO produces complex structural changes that could affect reliability and cyclability of ferroelectric devices. However, further insights from imaging buried films or studying domain dynamics through operando-compatible techniques like nano-focused X-ray imaging can increase our understanding of device performance. This may allow for better tailoring of electrode deposition to minimize domain reconfiguration or even strain engineering of ferroelectric domains through controlled top-layer deposition. Ultimately, understanding domain morphology and dynamics within buried ferroelectric stacks, which can differ substantially from bare films, will improve our ability to control and design next-generation computing and memory devices.

--------------

## Methods

**XRD measurements:** Diffraction measurements were conducted using the Diffraction Endstation of the NanoMAX beamline at MAX IV Laboratory. Symmetric scattering, as shown in Figure 1b, was used to probe the BFO $(003)_{pc}$ Bragg condition. Measurements were taken at 15 keV and an incident angle of 18°, producing a beam footprint of ~60 nm x 180 nm. The sample was mounted with the majority of ferroelectric stripes parallel to the beam direction (along DSO [001]) to maximize spatial resolution. Nano-XRD was performed, collecting high resolution 2D diffraction patterns of the $(003)_{pc}$ peak, as exampled in Figure 1c, at each probe position while scanning the sample under the X-ray beam with a 50 nm step size (example by the white dotted grid in Figure 1b). The center of mass of the Bragg peak was calculated for each probe position, producing a tilt map in Figure 1e. See Supplementary Note 1 for more details on COM and tilt determination.[22,23] Tilt maps shown here are produced from diffraction taken at a single angle instead of from a full 3D angular dataset, therefore they are only approximate (see Supplementary Note 3).

**Sample fabrication:** The BFO/SRO films were grown on single-crystalline$(110)_o$-oriented DSO substrates (CrysTec GmbH) by pulsed laser deposition using a 248 nm KrF excimer laser. The SRO buffer layer was deposited at 700°C under 0.016 mbar oxygen partial pressure with a laser fluence of 1.35 J cm$^{-2}$ and a laser repetition rate of 2 Hz. The BFO films were subsequently grown at 680°C under 0.12 mbar oxygen partial pressure with a laser fluence of 1.71 J cm$^{-2}$ and a laser repetition rate of 8 Hz.

After the cooling process, the films were transferred into the DC-magnetron sputtering chamber at a base pressure of ~$10^{-8}$ mbar. The top 100 nm Pt layer was deposited under an argon pressure of $10^{-3}$ mbar. The circular electrodes with 20 μm diameter were patterned by photolithography and argon plasma etching.

**Electrical poling:** The electrodes were biased with a home-built ferroelectric test system with a probe station and a PFM tip. The positive-up-negative-down (PUND) technique was used for two electrodes (e:6V, e:8V). A sequence of pulses was applied to the Pt top electrodes at 5 kHz. Firstly, a preset voltage pulse of negative polarity sets the polarization, and subsequently, two pulses of positive polarity were applied, followed by two pulses of opposing polarity. Electrode e2:6V was biased using a PFM tip at 2 Hz with -6V. The pulses were applied to the top electrode via the probe and PFM tip, and the bottom SRO electrode was grounded.

**PFM poling and measurements:** The PFM measurements and the electric-field tip poling were performed using a Bruker Multimode 8 atomic force microscope with μmasch HQ:NSC35/Pt tips in contact mode. During raster-scanning, a 2.5 V peak-to-peak AC voltage modulation was applied to the tip at 15 kHz. Ferroelectric poling was induced by applying a DC bias of ±5 V to the tip. The bottom SRO electrode was grounded.

**Acknowledgements:** This research was partially conducted at the NanoMAX beamline of MAX IV Laboratory under proposal numbers 20240018 and 20241230. Research conducted at MAX IV, a Swedish national user facility, is supported by Vetenskapsrådet (Swedish Research Council, VR) under contract 2018-07152, Vinnova (Swedish Governmental Agency for Innovation Systems) under contract 2018-04969 and Formas under contract 2019-02496. This project received funding from the European Research Council (ERC) under the European Union's Horizon 2020 research and innovation program (Grant 801847). Additional funding was provided by the Olle Engkvist Foundation, NanoLund, the Essence project and the Swedish Research Council (contract no. 2021-04273). M.T., I. E. and B.Y. acknowledge the ETH Zurich Research Grant funding under reference. 22-2 ETH-016 and the Swiss National Science Foundation under project no. 200021_188414 and 200021_231428.

**Conflict of interest:** Authors declare no financial/commercial conflicts of interest.

**TOC text:** Nanoprobe X-ray diffraction is used to image ferroelectric domains inside $BiFeO_3$-based capacitors, revealing striking differences from bare films such as local disorder in domain architecture. Sensitivity to ferroelectric reversal in poled capacitors is also demonstrated. Ultimately, previous assumptions about domain structure in buried ferroelectrics are challenged, making evident the need to further investigate domain morphology and dynamics within ferroelectric devices to control and design next-generation computing and memory devices.

**Supplementary Information:** Electrode modified domain morphology in ferroelectric capacitors revealed by X-ray microscopy

Megan O. Hill Landberg[1*], Bixin Yan[2], Huaiyu Chen[3], Efe Ipek[2], Morgan Trassin[2], and Jesper Wallentin[3]

**Supplemental Note 1:** 3D strain mapping analysis.

3D strain mapping was performed on an as-grown (AG) region of the film in the parallel scattering geometry. Scans were repeated at 6 angles (0.02° step size) to guarantee that the Bragg condition was fulfilled for the majority of scan positions, as seen in the integrated diffraction intensity map of a region within the AG film (Figure S1a). An example diffraction pattern for the AG region is shown in Figure S2a, the location of this pattern is marked by a red dot in Figure S2b. A rocking curve in the vicinity of this red dot is shown in Figure S2c. Example diffraction patterns along this rocking curve are also shown in Figure S2d.

Strictly speaking, precise strain and tilt mapping requires collection of 3D diffraction images as performed for this region. To quantify tilt and strain, first real space pixels, $\Delta_{x,y}$ on the detector image must be converted to reciprocal space units, $dq_1$ and $dq_2$, calculated as follows: $dq_{1,2} = \frac{2\pi \Delta_{x,y}}{D\lambda}$, where D is the detector distance from the sample and λ is the wavelength. The third dimension is defined by the angular steps taken (sample rotation steps $d\theta$): $dq_3 = \frac{2\pi \, d\theta \, \sin(\theta)}{\lambda}$ where θ is the Bragg angle. The reason that 3D diffraction patterns are required, is that the coordinates $q_1, q_2, q_3$ are not orthogonal, but $q_1$ and $q_3$ are dependent on each other. To convert to an orthogonal coordinate system, the following transformations were performed: $q_x = q_1 \cos(\theta)$; $q_y = q_2$; $q_z = q_3 - \sin(\theta)$.

Quantification of (003)pc strain and tilt were performed by extracting the peak centre-of-mass (COM) in $q_x$, $q_y$, and $q_z$. Given the high flux of the diffraction patterns (>10$^5$ integrated photons at each probe position) the COM can be resolved with sub-pixel resolution providing a tilt resolution better than 0.0044°. Relative tilts are calculated by measuring the deviation in the Bragg peak COM and converting to rotations around $q_z$ (α-tilt) and $q_y$ (β-tilt) as follows: $\alpha = \sin^{-1} \frac{q_y}{|Q|}$ and $\beta = \tan^{-1} \frac{q_x}{q_z}$, $|Q|$ is the modulus of all COM components. The d-spacing is defined by the modulus: $d = \frac{2\pi}{|Q|}$ and from the d-spacing a relative strain can be extracted $\epsilon_{(003)} = \frac{d - d_{ref}}{d_{ref}}$. Relative tilts and strain were calculated as compared to the mean value in the map.

Maps of the same AG region are shown for α, β, and (003) strain in Figure S1b, c, and d respectively. The striped domains are clearly observed in all images, with stripes visible in primarily the horizontal, but also the vertical direction. The average domain size, as extracted from the line cut in Figure S1e (marked by the dotted line in S1a), is 130 nm ± 10 nm, matching the expected spacing measured in PFM. Figure S1f schematizes relationship between α and β tilts with respect to the AG domain stripes. Unsurprisingly, the domain features are most pronounced in α as the ferroelastic domain stripes are formed due to the tilt mismatch between r$_4$(r$_2$) and r$_3$(r$_1$) variants.

The line cut shows the difference in α tilt between domains to be 0.015° ± 0.005° and a smaller tilt of <0.004° is identified along the domain stripes (β). For 71° striped domains in BFO, the domains are typically approximated to be fully coherent along the [010]pc direction but the observed tilt in β shows a tilt mismatch along this direction though it is ~25% of the [100] tilt. Beyond tilt, the ferroelastic domains present a strain variation of ~0.033%. Fig. S1e shows that strain is spatially offset from α tilt, this confirms expectations that strain comes from the tilt mismatch between r4 and r3, not from an inherent d-spacing difference between the two domain variants. Interestingly, β tilt is spatially aligned with strain (though reversed in magnitude).

The diffraction intensity in some regions of the film are lower, particularly in the regions between vertical and horizontal stripe domains (109° variant switches). In these regions, the Bragg condition is fulfilled (see Supplementary Note 5), however interference effects reduce the overall intensity. Additional interference effects may also be present due to 109° DWs, which would produce much higher frequency fringes, with oscillations larger than the Bragg peak width expected for 1-2 nm DWs.

While 3D strain mapping is more complete than 2D mapping, for large areas it is overly time consuming. Additionally, it was found that repetitive mapping of poled domain regions could act to depolarize the induced domain state, see Supplementary Note 6. However, repeated mapping of the AG domain region produced no noticeable changes in the domain structure.

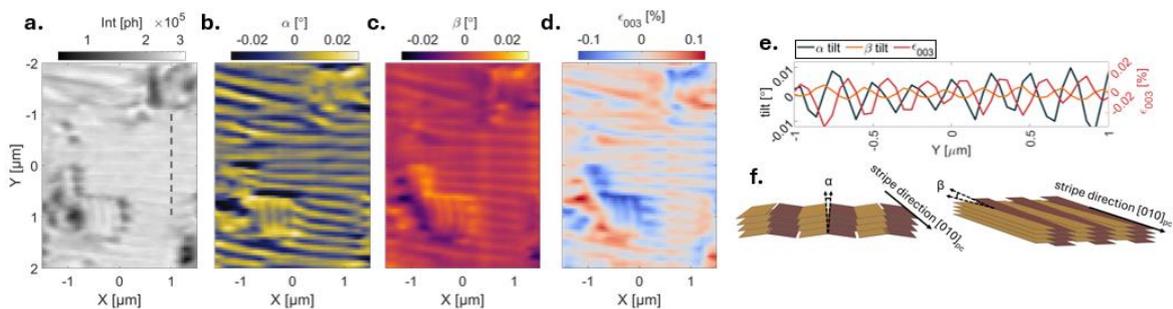

**Figure S1:** 3D strain mapping of AG region of the BFO film: (a) Integrated diffraction intensity, (b) α-tilt, (c) β-tilt, and (d) (003) strain. (e) Line cut extracted from dotted line in (a). (f) Schematic of α-tilt and β-tilt directions with respect to AG domains stripes.

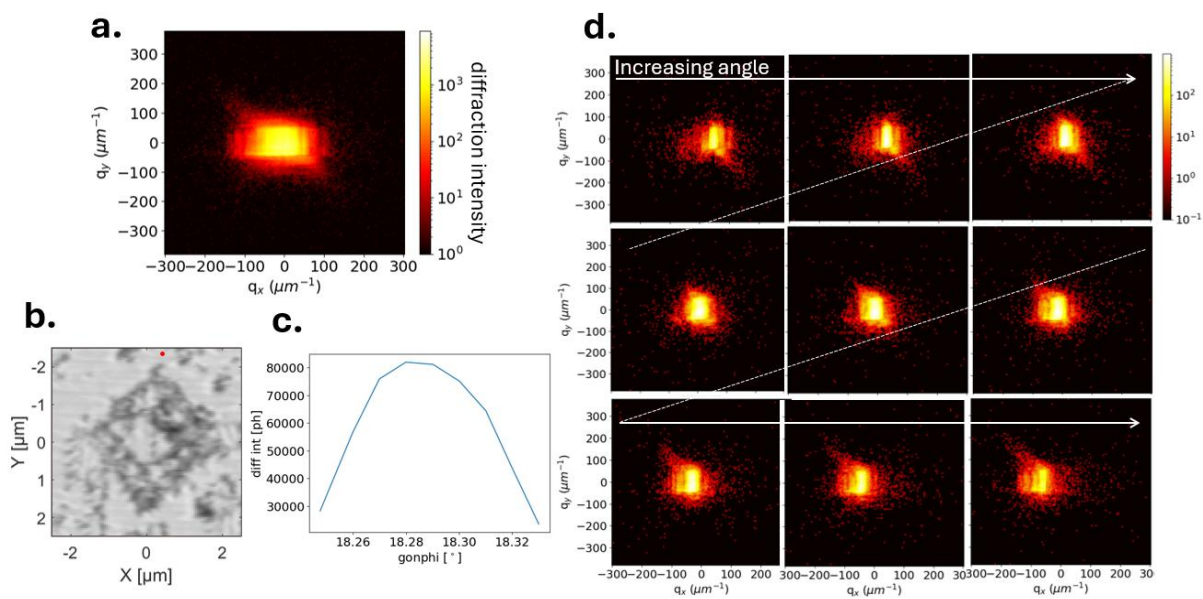

**Figure S2:** Example rocking curve from AG region of the BFO film. (a) Example Bragg peak extracted from the position marked by a red dot in (b). (c) Example rocking curve taken in the vicinity of the red dot in (b). (d) Extracted diffraction patterns along the rocking curve in (c).

**Supplemental Note 2:** Statistical analysis of nano-XRD tilt maps.

An additional BFO film, grown using the same conditions was measured with nano-XRD. Example maps for this second film are shown in Figure S3a with α-tilt and strain (004)pc shown on top and bottom respectively. Interestingly here, the ferroelastic domains are most pronounced when looking at the strain map. Here the striped domains are clearly visible in the AG region (-2.5 to 0 µm in Y) and a more disordered domain structure, with possibly more vertical stripes, is seen under the electrode (0 to 2.5 µm in Y).

A larger map was taken on this second BFO film to allow for statistical analysis of the real space α-tilt image. Figure S3b shows the larger area mapped and white dashed lines are used to outline the regions used for FFT analysis. These two regions were normalized and an FFT power spectrum was generated for each, the log power density is shown in Figure 1g. Figure S3b shows the normalized angular distribution of the power density. The AG regions shows primarily a sharp peak in the angular profile due to the strong ordering of the horizontal ferroelastic domains. However, for the electrode region, the power spread is much flatter and broader, this is indicative of a much more disordered domain structure present under the electrode. While FFT analysis is not possible on electrode:0V (Figure 1e) due to insufficient statistical area, it is possible to compare the distribution of α-tilt angles. This is shown as a histogram in Figure S3c. Here it is seen that the electrode region has a smaller spread of angles (narrower distribution) which may be related to the increased disorder in the electrode region, where the individual ferroelastic domains are not as well defined.

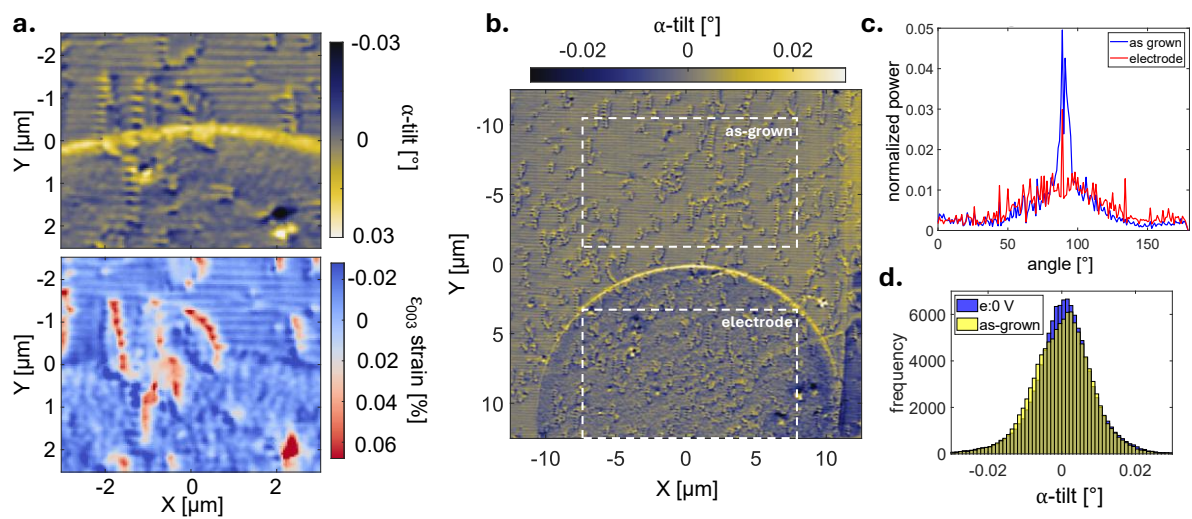

**Figure S3:** Additional analysis of domain tilts in AG and buried BFO. (a) Example tilt (top) and strain (bottom) maps of second BFO film (b) Large area mapped for a comparable statistics on the AG region and electrode region. White dotted lines show regions used for 2D FFT analysis. (c) Angular FFT profiles extracted from 2D FFT magnitude images in Figure 1g. (d) Histogram of α-tilt angular distribution for pristine electrode:0V compared to the surrounding AG region.

**Supplementary Note 3:** Determining strain without 3D rocking curves.

As described in Supplementary Note 1, true quantification of tilt and strain requires mapping diffraction patterns in 3D as strain is the modulus of the peak position in all three dimensions, $q_x, q_y, q_z$, and tilt has components in multiple q directions as well. However, instead of taking full 3D rocking curves (reciprocal space maps) at each position, a good approximation can be to find the angle at which the Bragg peak maximum occurs for each position, and then find the 2D COM on the detector to calculate the tilt and d-spacing change. In this approach, the COM of $q_1 (\sim q_x)$ corresponds to d-spacing change and $q_2 (q_y)$ to α-tilt. In other words, finding the θ angle (at each sample position) in which the diffraction intensity is maximized allows for the calculation of strain and α-tilt from a singular 2D diffraction pattern at each sample position.

In the case of mapping electrodes e:0V and e:6V in Figure 2, a rocking curve was taken in the uniform, AG region of the film, shown in Figure S2. The maximum angle was then used for the 2D mapping of e:0V and e:6V electrodes. While angular optimization was not done on the entire mapped region, it can be seen from the diffraction intensity maps (Figure 2d), that the scattering intensity does not vary significantly between the AG region and the electrode region – excluding regions in which further disorder occurs such as switching from horizontal to vertical domain stripes. This indicates that the Bragg condition is fulfilled both in the AG region and under the electrodes, within the angular step size used (0.05°). As such, it is appropriate to approximate 2D COM changes in $q_x$ as strain variations and $q_y$ as α-tilt variations.

**Supplementary Note 4:** Additional biased electrodes.

Figure S4 shows full maps of all electrodes for α-tilt (top) and $q_x$ (bottom). The most obvious changes, both for α and $q_x$, are observed for e:6V and e:8V. This includes a more pronounced higher α angle ring around the electrode and a larger difference in $q_x$ magnitude between the AG and biased regions. Electrode e2:6V looks more similar to the unbiased electrode e:0V, though there is a slightly stronger contrast in $q_x$. This is possibly because PUND measurements were not performed on e2:6V, it was instead polarized using direct bias from a PFM tip at a lower frequency (2 Hz) compared to the PUND biased e:6V and e:8V (5 kHz).

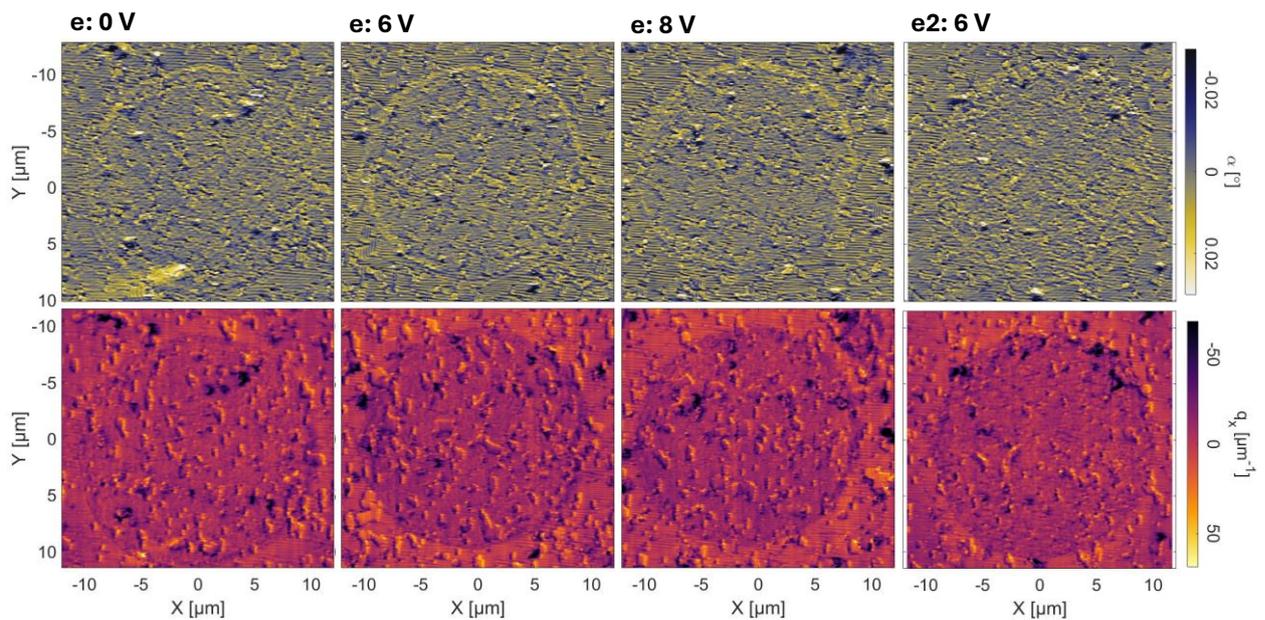

**Figure S4:** Full α-tilt (top) and $q_x$ (bottom) maps for all measured electrodes.

**Supplementary Note 5:** Interference effects due to domain walls.

Though diffraction intensity is lower in the regions between vertical and horizontal stripe domains (109° variant switches), the decreased intensity is likely from interference effects, opposed to being off of the Bragg condition. Figure S5a shows an example diffraction pattern from the 109° variant region, as marked by the blue dot in Figure S5b. Here we can observe interference effects, which could result from being in the vicinity of vertical domain stripes. Measuring a number of angles in the vicinity of this 109° variant, we cannot observe a clear rocking curve. However, looking at the individual diffraction patterns in Figure S5d, we do observe a mostly symmetric transition from fringes on the left of the peak to the right of the peak. Indeed, the Bragg maximum is defined not by the angle of maximum intensity, but instead as the angle at which the symmetric centre of the Bragg peak is observed (though these are usually equivalent). This indicates that the Bragg maximum has likely been fulfilled in this angular range despite the lack of clear intensity maximum. Additionally, particularly for the fourth pattern, we see further interference effects that could result from 109° DWs, which would produce much higher frequency fringes, resulting in the drop in intensity of the Bragg peak.

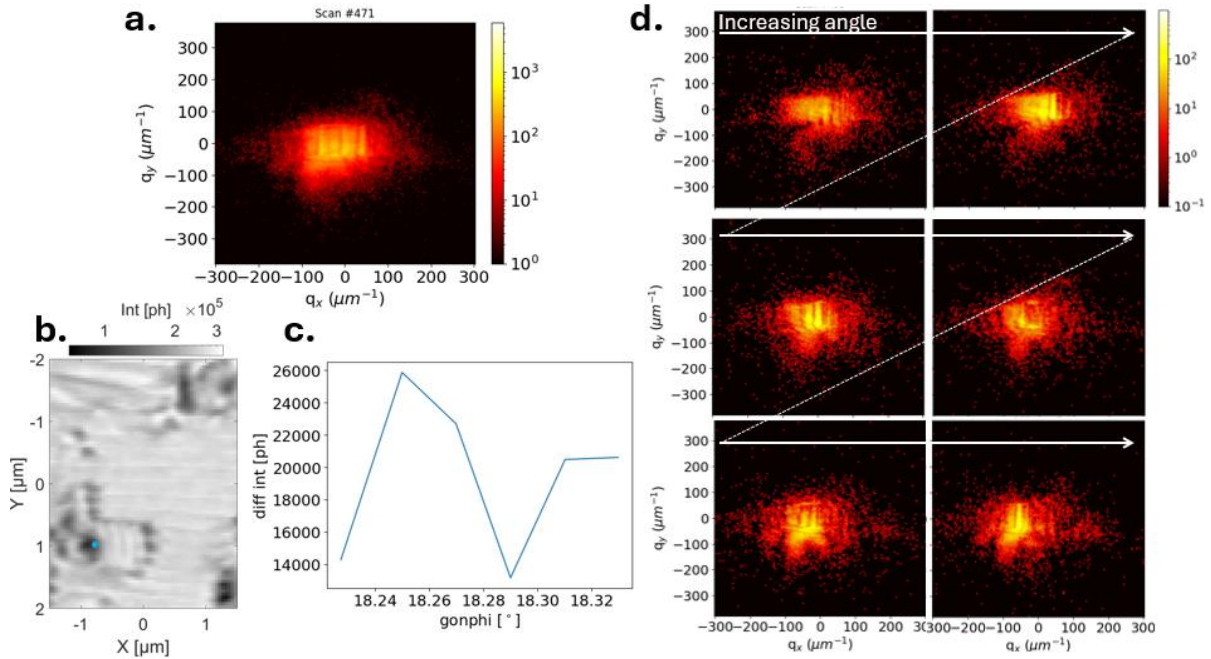

**Figure S5:** Example rocking curve from the 109° variant region of the BFO film. (a) Example Bragg peak extracted from the position marked by a blue dot in (b). (c) Example rocking curve taken in the vicinity of the blue dot in (b). (d) Extracted diffraction patterns along the rocking curve in (c).

The same interference effects are observed for the p-up polarized region of the box-in-box structure. Here a significantly lower intensity is observed as compared to the AG and the p-down regions. Figure S6a shows an example diffraction pattern, where significant interference is clearly observed as compared to the AG region in Figure S1a. This is taken in the p-up region marked by the red dot in Figure S6b. A rocking curve in the vicinity is shown in Figure S6c. Though there are some variations in intensity, a peak is observable, suggesting that here the Bragg condition is fulfilled, however interference effects result in the observed intensity reduction. For additional details, Figure S6d shows the diffraction patterns in the region across the rocking curve. These show a mostly symmetric effect with fringes transitioning from left to right of the primary peak with increasing angles.

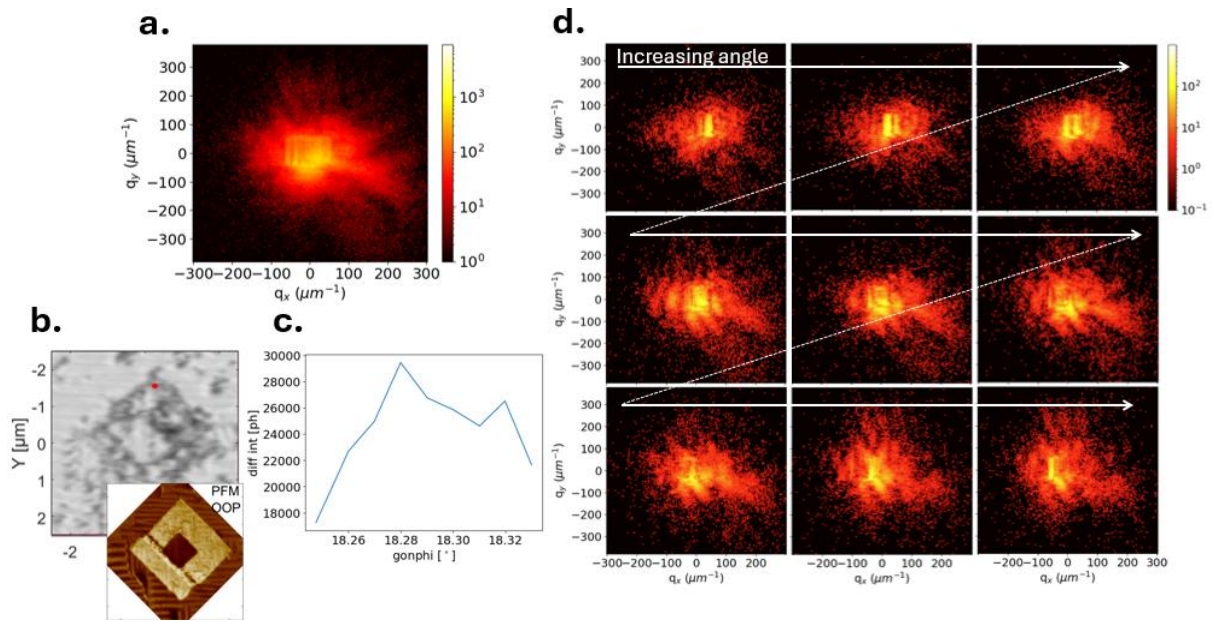

**Figure S6:** Example rocking curve from the p-up region of the box-in-box polarized region of BFO. (a) Example Bragg peak extracted from the position marked by a red dot in (b). (c) Example rocking curve taken in the vicinity of the red dot in (b). (d) Extracted diffraction patterns along the rocking curve in (c).

**Supplementary Note 6:** Additional box-in-box structure and beam damage testing.

An additional box-in-box region (2) was characterized and presented comparable results to the box-in-box 1 structure. Figure S7a shows the out-of-plane polarization measured by PFM. Results from the parallel and perpendicular scattering for box-in-box 2 are shown in (b) and (c) respectively. The top map shows the integrated diffraction intensity, center shows the $q_y$ COM, and lower shows $q_x$ COM.

After measuring the box-in-box region 4-5 times, the difference between the AG region and the polarized region is significantly reduced as seen by the maps in Figure S7d. This indicates that repetitive exposure to the X-ray beam can act to depolarize the film.

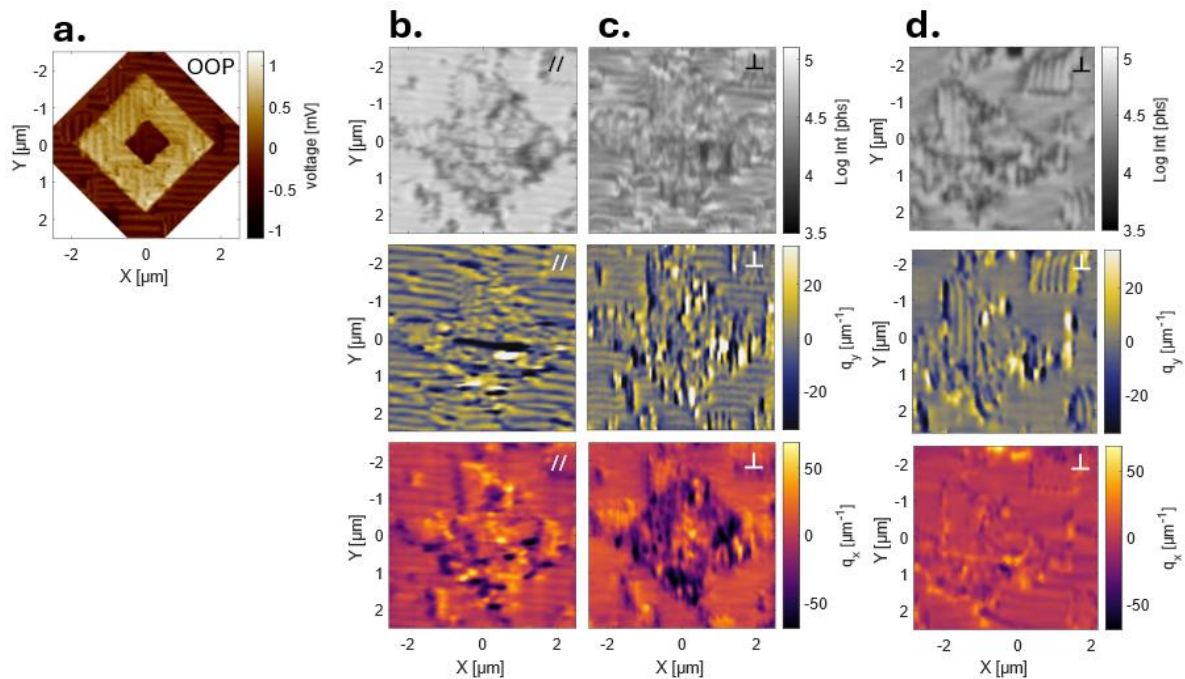

**Figure S7:** Box-in-box 2 structure and beam damage evaluation. (a) Out-of-plane PFM map of an additional box-in-box structure. (b) Parallel and (c) perpendicular scattering results. Top shows diffraction intensity, center shows $q_y$ COM, and bottom shows $q_x$ COM. (d) Map of box-in-box 2 after repeated measurements.

Though the X-ray beam appears to have an effect on the polarization of the film, it does not appear to easily damage the AG domain structure. This is shown by the repeated maps measured in the AG region of Figure S8.

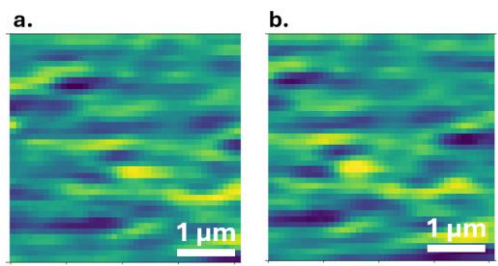

**Figure S8:** Beam damage test for AG region of BFO. Integrated diffraction intensity maps after a single exposure (a) and after 5 repetitions (b). The diffraction intensity is unchanged between the two maps despite a long exposure (0.1 s per point) and high flux (7e9 ph/s).